\documentclass[secnumarabic, graphics,floatfix, nofootinbib,tightenlines,nobibnotes, aps, prl, 12pt]{revtex4-2}
\usepackage[english]{babel}
\usepackage[utf8]{inputenc}
\usepackage{amsfonts}
\usepackage{multirow}
\usepackage{graphicx}
\usepackage{epstopdf}
\usepackage{hyperref}
\usepackage{latexsym}
\usepackage{amsmath}
\usepackage{amssymb}

\usepackage[utf8]{inputenc}
\hypersetup{
    colorlinks=true,
    linkcolor=red,
    citecolor=blue,
}
\usepackage{color}
\usepackage[T1]{fontenc}
\usepackage{txfonts}
\usepackage{mathrsfs}
\usepackage[center]{subfigure}

\begin{document}
 \setcounter{secnumdepth}{2}
 \newcommand{\bq}{\begin{equation}}
 \newcommand{\eq}{\end{equation}}
 \newcommand{\bqn}{\begin{eqnarray}}
 \newcommand{\eqn}{\end{eqnarray}}
 \newcommand{\nb}{\nonumber}
 \newcommand{\lb}{\label}
 
\title{On the late-time tails of massive perturbations in spherically symmetric black holes}

%\author{the authors}
\author{Wei-Liang Qian$^{2, 1, 3}$}\email[E-mail: ]{wlqian@usp.br}
\author{Kai Lin$^{4, 2}$}
\author{Cai-Ying Shao$^{5}$}
\author{Bin Wang$^{1, 6}$}
\author{Rui-Hong Yue$^{1}$}

\affiliation{$^{1}$ Center for Gravitation and Cosmology, School of Physical Science and Technology, Yangzhou University, 225009, Yangzhou, Jiangsu, China}
\affiliation{$^{2}$ Escola de Engenharia de Lorena, Universidade de S\~ao Paulo, 12602-810, Lorena, SP, Brazil}
\affiliation{$^{3}$ Faculdade de Engenharia de Guaratinguet\'a, Universidade Estadual Paulista, 12516-410, Guaratinguet\'a, SP, Brazil}
\affiliation{$^{4}$ Hubei Subsurface Multi-scale Imaging Key Laboratory, Institute of Geophysics and Geomatics, China University of Geosciences, 430074, Wuhan, Hubei, China}
\affiliation{$^{5}$ MOE Key Laboratory of Fundamental Physical Quantities Measurement, Hubei Key Laboratory of Gravitation and Quantum Physics, PGMF, and School of Physics, Huazhong University of Science and Technology, 430074, Wuhan, Hubei, China}
\affiliation{$^{6}$ School of Aeronautics and Astronautics, Shanghai Jiao Tong University, 200240, Shanghai, China}

%\date{\today}
\date{August 20th, 2022}

\begin{abstract}
It was first pointed out by Koyama and Tomimatsu that, under reasonable assumptions, the asymptotic late-time tails of massive scalar perturbations in the far zone of spherically symmetric black hole spacetimes decay universally as $t^{-5/6}$.
The late-time tail is furnished by the contribution from the branch cut of the frequency-domain Green's function, which is constructed in terms of two appropriate solutions of the corresponding homogeneous equation.
%In particular, the branch cut comes from the out-going waves that are subsequently backscattered at the spatial infinity at the low-frequency limit. 
%The in-going solution of the homogeneous equation, on the other hand, does not contain any singularity and subsequently plays a minor role.
%The present study is focused on some particular forms of the in-going wave that were considered unmanageable in the original derivations but nonetheless have been taken into account in some literature by other authors.
The present study focuses on some particular forms of the in-going wave that were not explicitly considered in the original derivations but nonetheless have been taken into account in the literature by other authors.
In this regard, we reassess the authors' arguments and provide a detailed complimentary analysis that covers a few specific aspects.
For some particular cases, the tail is found to possess the form $t^{-1}$.
We also discuss the possible implications of the present findings.

\end{abstract}

\maketitle

\newpage

\section{Introduction}\label{section1}

As one of the most intriguing concepts in theoretical physics, the black hole is an extreme manifestation of gravity.
The subject has received increasing attention in recent years owing to the advent of experimental detections of the gravitational wave, particularly those emanating from the coalescences of black hole binaries.
The gravitational radiations captured by the LIGO and Virgo collaboration~\cite{agr-LIGO-01, agr-LIGO-02, agr-LIGO-03, agr-LIGO-04} furnished direct evidence of the black holes.
Subsequently, it inaugurated a novel era where the predictions of General Relativity can be empirically tested against other possible alternatives in the strong-field regime. 

Besides the gravitational waves, the presence of a black hole can be inferred through the peculiar dynamics when it interacts with matter.
When a black hole is perturbed, the system oscillates dissipatively, and the evolution is characterized by the quasinormal modes (QNMs)~\cite{agr-qnm-review-02, agr-qnm-review-03, agr-qnm-review-06}.
The complex frequencies of such modes are found to be irrelevant to the initial conditions but governed by a few parameters of the black hole, where the number is subject to the constraint of various no-hair theorems~\cite{agr-bh-nohair-01, agr-bh-nohair-04}. 
On the theoretical side, the QNMs can be interpreted from various viewpoints.
The emergence of discrete complex frequencies is attributed to the in-going and out-going boundary conditions, defined for the non-Hamiltonian system at the horizon and outer spatial bound.
As a scattering problem, the QNMs correspond when the transmission or reflection coefficient becomes divergent.
Schutz and Will argued that it is feasible when the frequency, as a complex number, is located in the vicinity of the peak of the potential, which subsequently motivated the WKB approach~\cite{agr-qnm-WKB-01, agr-qnm-WKB-02}.
On the contrary, Ferrari and Mashhoon~\cite{agr-qnm-Poschl-Teller-02} pointed out that the scenario can be transformed into a bound state eigenvalue problem when the spatial coordinates of the wave function are extended to the imaginary axis by analytic continuation.
Reminiscent of the atomic spectrum of hydrogen in quantum mechanics, the continued fraction method was introduced by Leaver~\cite{agr-qnm-continued-fraction-01}.
Such an idea is also clearly demonstrated by the matrix method~\cite{agr-qnm-lq-matrix-01,agr-qnm-lq-matrix-02,agr-qnm-lq-matrix-03,agr-qnm-lq-matrix-04} proposed by some of us, where one solves a matrix equation for the complex eigenvalues.
By replacing the spacelike infinity with a null infinity $\mathscr{I}^+$, the hyperboloidal approach~\cite{agr-qnm-hyperboloidal-01} also successfully reformulates the problem into an eigenvalue problem of a non-selfadjoint operator.
The latter has been utlized in the recent developments on QNM instability~\cite{agr-qnm-instability-07, agr-qnm-instability-13, agr-qnm-instability-14}.
In terms of Green's function, Leaver presented the problem in terms of the Fourier spectrum decomposition~\cite{agr-qnm-21}, which was in turn reformulated by Nollert and Schmidt~\cite{agr-qnm-29} using Laplace transform.
For these approaches, the QNMs are associated with the poles of the frequency-domain Green's function.

A late-time tail often features the last stage of quasinormal oscillations.
Besides numerical approaches, the relevant time profile can also be analyzed using Green's function.
To be specific, the QNMs correspond to the poles, while the branch cuts govern the late-time tails on the lower half plane\footnote{In this study, we adopt the notion of Fourier transform instead of Laplace one.} of Green's function.
As first pointed out by Price, the late-time tails largely follow an inverse power-law form~\cite{agr-qnm-tail-01}. 
Mathematically, the origin of the late-time tail is attributed to the branch cut and is largely understood as a manifestation that the spectrum decomposition in terms of the QNM poles is not complete.
For massless perturbations, the tail is due to a branch cut placed on the negative imaginary axis of the frequency.
The brach cut in question stretches out from the origin, which is closely connected with the fact that the tail is a late-time phenomenon.
Moreover, it is governed by the asymptotic properties of the potential, which subsequently invited the interpretation in terms of the backscattering of the wave packets by the spacetime far away from the horizon~\cite{agr-qnm-tail-05}.
For massless scalar perturbations in the Schwarzschild black hole metric, the tail was found to be ${t^{ - (2l + 3)}}$ or ${t^{ - (2l + 2)}}$~\cite{agr-qnm-tail-01}.
Further studies for various different metrics~\cite{agr-qnm-tail-05, agr-qnm-tail-06, agr-qnm-tail-09, agr-qnm-tail-11, agr-qnm-tail-12} strongly indicated that such an inverse power-law form is a general feature for massless perturbations in spherical spacetimes.
Among others, a comprehensive study was performed by Ching {\it et al.}~\cite{agr-qnm-tail-05, agr-qnm-tail-06}.
Besides the power-law form, the exponential tails have also been observed in asymptotically de Sitter spacetimes~\cite{agr-qnm-tail-07, agr-qnm-tail-08}.
For massive perturbations, on the other hand, Hod and Piran~\cite{agr-qnm-tail-16} discovered that an intermediate late-time tail in Reissner-Nordstr\"om spacetime follows the form ${t^{ - (l + 3/2)}}\sin (\mu t)$, where $\mu $ is the mass of the field.
In a series of seminal papers, Koyama and Tomimatsu investigated the late-time tails in the Reissner-Nordstr\"om~\cite{agr-qnm-tail-20} and Schwarzschild~\cite{agr-qnm-tail-20, agr-qnm-tail-21} spacetimes and found it possess the asymptotic form ${t^{ - 5/6}}\sin (\mu t+\varphi)$.
Furthermore, they demonstrated that the obtained result is essentially a universal feature for spherically symmetric black holes, under moderate assumptions~\cite{agr-qnm-tail-22}.
The analytic result was confirmed numerically for different angular momenta and various types of perturbations~\cite{agr-qnm-tail-24, agr-qnm-tail-35, agr-qnm-tail-36}.
In all the above cases, there are two branch points located on the real axis of the frequency, which are connected by a branch cut conveniently chosen to lie on the real axis.
More recently, Cardoso {\it et al.} pointed out~\cite{agr-qnm-echoes-01} that the late-stage ringdown may serve to probe the properties of the horizon and discriminate between different gravitational systems.
Following this line, the studies of echoes in various systems, and inclusively, the exotic compact objects such as gravastar, boson star, and wormhole.
From an empirical perspective, both the echoes and tails occur in the last stage of the time profile.
However, mathematically, the echoes are mainly attributed to a novel spectrum of poles of the Green's function~\cite{agr-qnm-echoes-15, agr-qnm-echoes-16, agr-qnm-lq-03, agr-qnm-echoes-20}, that is distinct from those of the late-time tail.
These intriguing characteristics further broaden the underlying physics regarding the late-time waveforms.

The present work involves a study of the universal properties of the late-time tails of massive perturbations in spherically symmetric black holes.
We revisit the general derivation given by Koyama and Tomimatsu~\cite{agr-qnm-tail-22} and discuss possible modifications when applied to some particular cases.
To be more specific, the late-time tail is furnished by the contributions originating from the branch cut of the out-going wave function of the corresponding homogeneous equation.
We explicitly show that the branch cut in question resides below the real axis of the frequency complex plane due to the physical boundary condition.
On the other hand, the in-going solution of the homogeneous equation does not contain any singularity and is mainly understood to play a minor role.
However, we point out that the original derivation has to be adapted when the in-going waveform contains only one of the two Whittaker functions $M_{\kappa, \lambda}$ and $M_{\kappa, -\lambda}$.
Such a choice cannot be entirely ruled out because its context might be physically relevant, as considered in the literature by some authors~\cite{agr-qnm-tail-31, agr-qnm-tail-32, agr-qnm-tail-33, agr-qnm-tail-34, agr-qnm-tail-37}.
Compared to the latter, a crucial point of our approach resides in the requirement that the relevant waveform does not contain any singularity. 
As elaborated further discussed in Appendix~\ref{indxA}, it is a physically relevant requirement, which is also manifestedly consistent with the present mathematical formalism.
Our derivations indicate complementarily that an asymptotic power-law form holds as $t^{ - 1}\sin(\mu t)$ for some apecific cases.

The remainder of the paper is organized as follows.
In the next section, we present the general form of the spherically symmetric black hole metric and the approximation that gives rise to the radial master equation of massive perturbations.
The asymptotic properties of the late-time tails are studied in Sec.~\ref{section3} by utilizing the analytic properties of the Whittaker functions and the boundary conditions.
We analyze the location of the branch cut and consider two specific cases that complement the scenario to which the original formulae cannot be straightforwardly applied.
Further discussions regarding relevant studies in the literature and concluding remarks are given in Sec.~\ref{section4}.

\section{Green's function for massive scalar perturbations in spherically symmetric black hole spacetimes}\label{section2}

Following~\cite{agr-qnm-tail-22} one considers a static spherically symmetric black hole metric given by 
\begin{eqnarray} \label{SSSmetric}
  ds^2=-f(r)dt^2+h(r)dr^2+r^2(d\theta ^2+\sin ^2\theta d\varphi ^2) ,
\end{eqnarray}
where, in the present study, the observation takes place in the far region, and one considers the following expansions of $f$ and $h$ as a power series in $1/r$
\begin{eqnarray} \label{eqf}
  f= 1-\frac {2M}{r}+\frac {Q^2}{r^2}+O\left(\frac{1}{r^3}\right)
\end{eqnarray}
and
\begin{eqnarray} \label{eqh}
  h= 1+\frac {2M'}{r}+\frac {Q^{'2}}{r^2}+O\left(\frac{1}{r^3}\right) .
\end{eqnarray}
Besides the gravitational mass $M$, the metric is specified up to second-order terms, characterized by the parameters $M'$, $Q$, and $Q'$.
It is noted that the proof does not reside in that the metric Eq.~\eqref{SSSmetric} is a solution of Einstein field equations.
However, it is assumed that the conclusion drawn from the analysis based on the second-order expansions Eqs.~\eqref{eqf} and~\eqref{eqh} is robust.
As will become apparent below, such a treatment essentially evaluates the dominant contributions from the leading-order term, which contains a branch cut.
A similar approach has also been employed for massless perturbations in the literature~\cite{agr-qnm-tail-05, agr-qnm-tail-06} and the results were shown to be valid for different metrics and various types of perturbations.

Massive scalar perturbations are governed by the Klein-Gordon equation
\begin{eqnarray}
\frac{1}{\sqrt{-g}}\partial_{\mu}(\sqrt{-g}g^{\mu\nu}\partial_{\nu}\Phi(t,r,\theta,\varphi))=\mu^2 \Phi(t,r,\theta,\varphi) ,\label{eq5}
\end{eqnarray}
where $\mu$ is the mass of the field $\Phi$.
To proceed, one uses the standard method of separation of variables for $\Phi(t,r,\theta,\varphi)$, namely,
\begin{equation}
\Phi(t,r,\theta,\varphi)=\sum_{\ell,m}\frac{\Psi(t,r)}{r}Y_{\ell m}(\theta,\varphi)\label{eq4},
\end{equation}
where $Y_{\ell m}(\theta,\varphi)$ is the spherical harmonics and $\ell$ and $m$ stand for the angular and azimuthal number, respectively.
The resultant radial master equation for a given multiple moment $\ell$ reads
\begin{eqnarray}
\label{masterEq}
\left[\frac{\partial ^2}{\partial t^2}  - \frac{\partial ^2}{\partial r_{\ast}^2} +V(r) \right]\Psi (t,r) =0,
\end{eqnarray}
where $r_{\ast}$ is the tortoise coordinate
\begin{eqnarray}
  \frac{dr_{\ast}}{dr} &=& \sqrt{\frac{h}{f}},
\end{eqnarray}
and $V$ is the effective potential
\begin{eqnarray} \lb{effV}
  V=f\left[\frac{1}{r\sqrt{fh}}\left(\sqrt{\frac{f}{h}}\right)' +\frac{\ell(\ell+1)}{r^2}+\mu^2 \right] ,
\end{eqnarray}
where the prime denotes a derivative with respect to the areal radius $r$.
It is noted that the radial wave function does not depend on the azimuthal number $m$, owing to the spherical symmetry.
For simplicity, we have also omitted the index $\ell$ in $\Psi(t,r)$.

In terms of the Green's function~\cite{book-methods-mathematical-physics-04}, the solution of Eq.~\eqref{masterEq} is given by
\begin{eqnarray}
  \Psi (r_{\ast},t) &=&\int [G(r_{\ast},r_{\ast}';t)\psi_t(r_{\ast}',0) +G_t(r_{\ast},r_{\ast}';t)\psi(r_{\ast}',0)] dr_{\ast}'
\end{eqnarray}
for $t\ge 0$, where the retarded Green's function $G$ satisfies
\begin{eqnarray}\lb{defGreen}
  \left[\frac{\partial ^2}{\partial t^2} -\frac{\partial ^2}{\partial r_{\ast}^2} +V\right]G(r_{\ast},r_{\ast}';t)&=&\delta(t)\delta(r_{\ast}-r_{\ast}') .
\end{eqnarray}

Its Fourier transform,
\begin{eqnarray}
  \tilde{G}(r_{\ast},r_{\ast}';\omega) &=&\int G(r_{\ast},r_{\ast}';t)e^{i\omega t}dt ,
\end{eqnarray}
can be expressed in terms of two linearly independent solutions, denoted by $\tilde\Psi_i$, where $i=1, 2$, for the homogeneous equation~\cite{agr-qnm-review-02}
\begin{eqnarray}
\label{mode-radial}
\left[\frac{\partial ^2}{\partial r_{\ast}^2} +\omega ^2 -V \right]\tilde{\Psi}=0 .
\end{eqnarray}
To be specific, $\tilde{G}(r_{\ast},r_{\ast}';\omega)$ can be given by
\begin{equation}\lb{GFomega}
  \tilde{G}(r_{\ast},r_{\ast}';\omega) =\frac{1}{W}
\left\{
  \begin{array}{l@{\quad \quad}l}
\tilde{\Psi} _1(r'_{\ast},\omega)\tilde{\Psi} _2(r_{\ast},\omega) &r_{\ast}>r'_{\ast}\\
\tilde{\Psi} _1(r_{\ast},\omega)\tilde{\Psi} _2(r'_{\ast},\omega)  &r_{\ast}<r'_{\ast}
  \end{array} 
\right. ,
\end{equation}
where $W$ is the Wronskian
\begin{eqnarray}\lb{defWronskian}
  W \equiv W(\tilde\Psi_1, \tilde\Psi_2) = \tilde{\Psi} _1\tilde{\Psi} _{2,r_{\ast}} -\tilde{\Psi} _{1,r_{\ast}}\tilde{\Psi} _2.
\end{eqnarray}
The boundary conditions for the solution $\tilde{\Psi}_1 $ is that it should be an in-going wave on the event horizon~\cite{agr-qnm-review-02}.
Moreover, it is noted that $\tilde{\Psi}_1 $ is well-behaved as it does not contain any branch cut~\cite{agr-qnm-tail-06, agr-qnm-tail-16, agr-qnm-tail-13, agr-qnm-tail-22}.
On the other hand, $\tilde{\Psi}_2 $ is required to be an out-going wave at spatial infinity.
Its branch cut eventually gives rise to the late-time tail, as explored in the following sections.

Using the specific forms, Eqs.~\eqref{eqf} and~\eqref{eqh}, one may expand the master equation Eq.~\eqref{masterEq} and neglect terms of order $O(1/r^2)$ and higher to find
\begin{eqnarray}
\label{masterEq-Whittaker}
%  \frac{B}{A}\partial _{t}^2\Psi +
\frac{\partial ^2\tilde{\Psi}}{\partial r^2} -U \tilde{\Psi}=0,
\end{eqnarray}
where %why the first-order derivative $\frac{M+M'}{r^2}\frac{\partial \tilde{\Psi}}{\partial r}$ is thrown away?
\begin{eqnarray}
\label{effU}
  U %&=& -\frac{B}{A}\omega^2 + V \nb\\
  &=&(\mu^2-\omega ^2) 
%-\frac{2M\omega ^2}{r}+\frac{2M'(\mu^2-\omega ^2)}{r}
-\frac{2(M+M')\omega ^2-2M'\mu^2}{r} 
%-\frac{\lambda ^2-\frac 14}{r^2} ,
-\frac{4(M^2+M M')\omega^2-(Q^2-{Q'}^2)\omega^2-{Q'}^2\mu^2-\ell(\ell+1)}{r^2} .\nb\\
\end{eqnarray}

The above approximate form is readily recognized to be Whittaker differential equation by introducing the variable
\begin{eqnarray}\lb{defX}
  x&=& 2\varpi r, 
\end{eqnarray}
where
\begin{eqnarray}\lb{defVP}
  \varpi =\sqrt{\mu^2-\omega ^2}.
\end{eqnarray}
To be specific, one arrives at the standard form~\cite{book-methods-mathematical-physics-05}
\begin{eqnarray}
\label{master-Whittaker}
\frac{d^2\tilde{\Psi}}{dx^2}+\left[-\frac 14 +\frac{\kappa}{x}
-\frac{\lambda ^2-1/4}{x^2}
\right]\tilde{\Psi}=0,
\end{eqnarray}  
where the coefficients $\kappa$ and $\lambda$ are given by
\begin{eqnarray}\lb{CoKappa}
  \kappa =\frac{M\mu^2}{\varpi}-(M+M')\varpi ,
\end{eqnarray}
and
\begin{eqnarray}\lb{CoLambda}
  \lambda =\sqrt{(Q^2-{Q'}^2)(\mu^2-\varpi^2)-4(M^2+{M}M')(\mu^2-\varpi^2)+{Q'}^2\mu^2+\left(\ell+\frac12\right)^2} .
\end{eqnarray}
A few comments are in order. 
The master equation Eq.~\eqref{master-Whittaker} can be viewed as a feasible approximation mainly for the far region.
In other words, in the vicinity of the horizon, the properties of the effective potential Eq.~\eqref{effU} no longer adequately reflect those of Eq.~\eqref{effV}.
As a result, the in-going wave boundary condition of $\tilde\Psi_1$ is likely to be distorted so that the feature does not necessarily manifest itself in a solution in the far region.
Usually, some ``matching'' process~\cite{agr-qnm-tail-20, agr-qnm-tail-21} should be employed if the specific waveform is of interest.
Fortunately, reminiscent of the case of massless perturbations~\cite{agr-qnm-tail-06}, for most of the time, it turns out~\cite{agr-qnm-tail-22} that the particular form of $\tilde\Psi_1$ is not a crucial ingredient for the late-time tails.
In this study, however, we will also explore a few exceptions in Sec.~\ref{subsec1} and~\ref{subsec2}.
Meanwhile, what remains essential is that $\tilde{\Psi}_1 $ shall {\it not} contain any branch cut, as it is a physically pertinent requirement for a black hole metric.

\section{Universal asymptotic properties of late-time tails} \lb{section3}

One may investigate the properties of the late-time tails in terms of the time-domain Green's function, which can be obtained through the inverse Fourier transform
\begin{eqnarray}
\label{time-Green-contour}
  G(r, r';t)&=& \frac{1}{2\pi}\int _{-\infty }^{\infty } \tilde{G}(r, r';\omega)e^{-i\omega t}d\omega ,
\end{eqnarray}
where $\tilde{G}(r, r';\omega)$ is given by Eq.~\eqref{GFomega}.
The solutions $\tilde{\Psi}_{1,2}$ are proper linear combinations of the Whittaker functions~\cite{book-methods-mathematical-physics-05}
\begin{eqnarray}\lb{WF2}
M_{\kappa,\lambda}\left(x\right) &=& \exp\left(-x/2\right)x^{\lambda+\tfrac{1}{2}}M\left(\lambda-\kappa+\frac{1}{2}, 1+2\lambda, x\right), \nb\\
W_{\kappa,\lambda}\left(x\right) &=& \exp\left(-x/2\right)x^{\lambda+\tfrac{1}{2}}U\left(\lambda-\kappa+\frac{1}{2}, 1+2\lambda, x\right),
\end{eqnarray}
where $M\left(\lambda-\kappa+1/2, 1+2\lambda, x\right)$ and $U\left(\lambda-\kappa+1/2, 1+2\lambda, x\right)$ are confluent hypergeometric functions of the first and second kinds. 
For given $\kappa$ and $\lambda$, one may choose either $\{ M_{\kappa,\lambda}, M_{\kappa,-\lambda} \}$ or $\{ M_{\kappa,\lambda}, W_{\kappa,\lambda} \}$ as the two linearly independent solutions.
According to Jordan's lemma, for an observer in the far region $t > r - r' > 0$ one may enclose the contour of the integral in Eq.~\eqref{time-Green-contour} by the lower half of the complex plane.
Here, the poles associated with the Wronskian on the denominator is not relevant, since we are only interested in the branch cut of $\tilde\Psi_2$.
In particular, if some branch cut exists but were located on the upper half of the frequency plane, it will not contribute to the late-time tail in the retarded Green's function.
Therefore, one must also justify that not only $\tilde\Psi_2$ contains some branch cut but it sits below the real axis of $\omega$, as well.
The appropriate choice is
\begin{eqnarray}
\label{psi2form}
\tilde{\Psi}_2(\varpi ,r)&=&W_{\kappa , \lambda}(2\varpi r) .
\end{eqnarray} 

The above statement can be justified as follows.
In order to apply any theorem regarding contour integral, the analytic continuation of $\tilde\Psi_2$ on the complex $\omega$ plane can be carried out, through the intermediate variable $\varpi$, from its definition on the real axis.
From the physical viewpoint, $\tilde\Psi_2$ is also subject to the appropriate asymptotic behavior as $r\to +\infty$, for a given $\omega$ on the real axis, where the frequency-domain Green's function Eq.~\eqref{GFomega} is defined.
To be specific, it must asymptotically converge to an out-going wave travelling along the positive direction of $r$-axis when $|\omega| > \mu $, namely, $\tilde\Psi_2\sim \exp\left(-x/2\right) \sim \exp\left(i |\varpi| r_* ~ \mathrm{sgn}\omega\right)$.
On the other hand, it is suppressed exponentially when $|\omega| < \mu $, namely, $\tilde\Psi_2\sim \exp\left(- |\varpi| r_* \right)$.
In other words, the argument of Eq.~\eqref{defVP} on the real axis is defined to be
\begin{eqnarray}\lb{argVP}
  \varpi
&=&    
\left\{
  \begin{array}{l@{\quad \quad}l}
+i\left|\sqrt{\omega ^2-\mu^2}\right|& \omega < -\mu \\
\left|\sqrt{\mu^2-\omega ^2}\right| & |\omega| <\mu\\
-i\left|\sqrt{\omega ^2-\mu^2}\right| & \omega >\mu 
  \end{array}
\right. ,
\end{eqnarray}
It is straightforward to show that Eq.~\eqref{argVP} is indeed satisfied by an analytic function, as will be elaborated below.
Indeed, by defining the argument of $\varpi$ on the real axis of $\omega$, Eq.~\eqref{argVP} removes the arbitrariness regarding how $\omega$ bypasses the branch points, and subsequently, the analytic continuation of $\tilde\Psi_2$ into the complex plane.
Moreover, the argument of $\varpi$ must be consistent with a second constraint that, the proper choice among the linear combinations of Whittaker functions that provides an asymptotic out-going wave, $W_{\kappa,\lambda}$, is valid only when $|\arg x | = |\arg \varpi | < 3\pi/2$~\cite{book-methods-mathematical-physics-07}.
As a result, these conditions tightly dictate the argument of $\varpi$ as $\omega$ moves along the real axis from $-\infty$ to $\infty$.
To be more specific, $\arg\varpi$ decreases continuously from $\pi/2$ to $0$ as $\omega$ circles from the above in the immediate vicinity of the branch point $\omega=-\mu$, and then continue to decrease from $0$ to $-\pi/2$ as it likewise bypasses the second branch point at $+\mu$.
We note that other possible intervals for $\arg x$, such as $3\pi/2 \le \arg\varpi\le 5\pi/2$ or $-5\pi/2 \le \arg\varpi\le -3\pi/2$, must be excluded as they invalidate the asymptotic behavior.
The above choice maps the argument from $[\pi, 0]$ of $\omega$ to $[\pi/2, -\pi/2]$ of $\varpi$.
In other words, if one tries to introduce an alternative contour that bypasses at least one of the branch points below the real axis, we can show that Eq.~\eqref{argVP} will not be satisfied.
As one considers the contour that goes alone the real axis of $\omega$ while bypassing both branch points from the above, Eq.~\eqref{argVP} can be rewritten into the following form 
\begin{eqnarray}\lb{defContourVP}
%\varpi =e^{-i\frac{\pi}{2}}\left[(\omega-\mu)(\omega -(-\mu))\right]^{\frac12}.
\varpi =e^{-i\frac{\pi}{2}}\left(\omega^2-\mu^2\right)^{\frac12} ,
\end{eqnarray}
where
\begin{eqnarray}\lb{defArgVP}
0 \le \arg(\omega-\mu), \arg(\omega+\mu) \le \pi , \nb
\end{eqnarray}
which is manifestly analytic above the contour in question.
Together with the asymptotic expansion of $W_{\kappa,\lambda}$, one ascertains that $\tilde\Psi_2$ is indeed analytic on the upper half plane of $\omega$.
Now, for an arbitrary analytic function $g$, $g(\varpi)$ as a function of $\omega$ usually possesses two branch points at $\omega=\pm \mu$.
Due to physical boundary conditions, the above considerations show that the branch cut that joins the two branch points must be placed on the lower half-plane.
Indeed, this conclusion is also expected based on heuristic arguments, and we will come back to further discussions in the last section of the paper.

\begin{figure*}[htbp]
\centering
\includegraphics[scale=0.45]{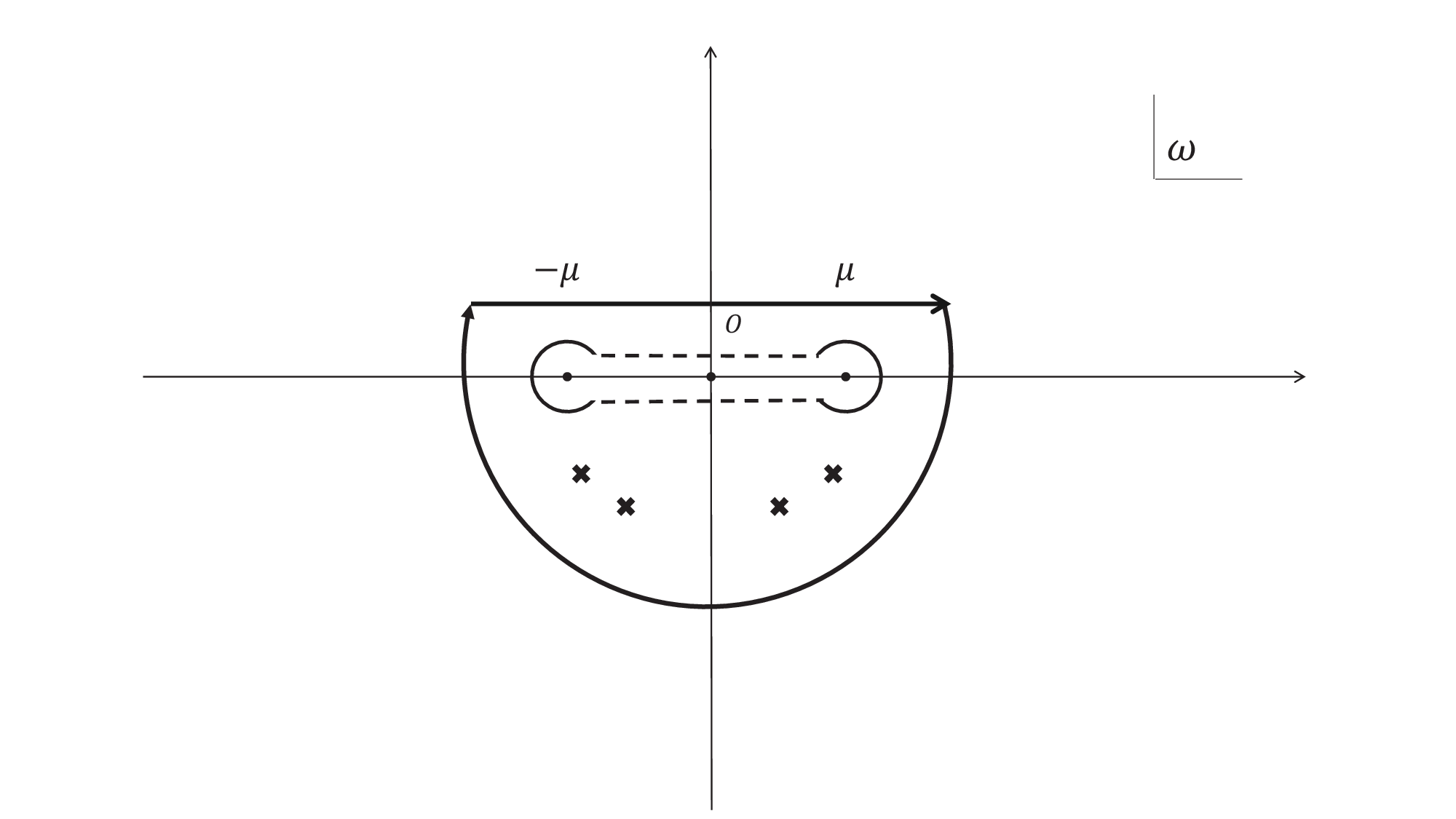}
\caption{The chosen contour of the integration and the branch cut on the complex plane of the frequency-domain Green's function.
The inverse Fourier transform corresponds to the integral along the real axis.
It is complemented by an integral along a large semicircle in the lower half of the complex plane, which is subsequently deformed to go around the quasinormal poles and branch cut indicated in the plot.}
\label{Fig1}
\end{figure*}

Different from the case of massless perturbations, for which the branch cut is typically caused by a function of the form $\omega^{1/2}$, it is essential to note that the infinity is no longer a branch point of $g(\varpi)$.
Therefore, as one closes the contour integral of $\omega$ counter-clockwise alone a large circle in the entire complex plane, the argument of $\varpi$ can be consistently chosen to lag behind that of $\omega$ by $\pi/2$.
It thus varies continuously within the interval $[-\pi, \pi]$, inside the range where $g=W_{\kappa,\lambda}$ is analytic. 
Subsequently, the contribution of the counter integral of Eq.~\eqref{time-Green-contour} comes entirely from the branch cut associated with $\varpi$. 
A convenient choice is to place the branch cut parallel to and immediately below the real axis of $\omega$~\cite{agr-qnm-tail-16, agr-qnm-tail-22} as shown in Fig.~\ref{Fig1}.
Thus we have, anywhere away from the branch cut,
\begin{eqnarray}\lb{defArgVPall}
-\pi &\le& \arg(\omega-\mu) \le \pi , \nb \\
0 &\le& \arg(\omega+\mu) \le 2\pi . \nb
\end{eqnarray}
As a result, the relevant contribution is governed by the difference between the values of the integrand of Eq.~\eqref{time-Green-contour} on the two opposite sides of the branch cut.
The argument of either factor must evolve while avoiding the chosen branch cut while encircling either one of the branch points.
It is easy to ascertain that the variable $\varpi$ is effectively differed by the transformation $\varpi \to \varpi e^{i\pi}$, thus Eq.~\eqref{time-Green-contour} gives
\begin{eqnarray}
\label{Fvarpi-contour}
  G(r, r';t)&=& -\frac{1}{2\pi}\int _{-\mu }^{\mu } F(r, r';\omega)e^{-i\omega t}d\omega ,
\end{eqnarray}
where
\begin{equation} \lb{defF}
F (\varpi )\equiv \frac{\tilde\Psi_1\left( r', \varpi e^{i\pi} \right)\tilde\Psi_2\left( r, \varpi e^{i\pi }\right)}{W\left( \varpi e^{i\pi} \right)}
- \frac{\tilde\Psi_1\left( r', \varpi \right) \tilde\Psi_2\left( r, \varpi \right)}{W(\varpi )}  .
\end{equation}

For $\tilde\Psi_1$, one requires that the wave form does not contain any discontinuity when crossing the branch cut.
The most general form that satisfies the requirement by combining the Whittaker functions is
\begin{eqnarray}
\label{psi1form}
\tilde{\Psi}_1(\varpi ,r)&=&A(\varpi) M_{\kappa , \lambda}(2\varpi r)+B(\varpi) M_{\kappa , -\lambda}(2\varpi r),
\end{eqnarray} 
where 
\begin{eqnarray}
  A=A(\varpi)= a\varpi ^{-1/2-\lambda}
\end{eqnarray}
and
\begin{eqnarray}
  B=B(\varpi) = b\varpi ^{-1/2+\lambda},
\end{eqnarray} 
where $a$ and $b$ are some one-valued even functions of $\varpi$.
Further discussions regarding the analytic properties of $\tilde{\Psi}_1$ are given in Appendix~\ref{indxA}.
As discussed above, such a waveform usually contains both the in-going and out-coming components.

By using Eqs.~\eqref{psi2form} and~\eqref{psi1form}, one can estimate the late-time asymptotic form of Eq.~\eqref{time-Green-contour}.
In what follows, we give an account for the discontinuity of the relevant quantities across the branch cut under the transformation $\varpi\to \varpi e^{i\pi}$.
Subsequently, we elaborate on the specific case for evaluating the Wronskian and the integration by saddle point approximation.

By making use of the relations~\cite{book-methods-mathematical-physics-06}
\begin{equation}\label{N36}
W_{\kappa, \lambda}(2\varpi r) = \frac{\Gamma ( -2\lambda)}{\Gamma \left( - \lambda - \kappa+\frac12 \right)}M_{\kappa, \lambda}(2\varpi r) + \frac{\Gamma (2\lambda)}{\Gamma \left(\lambda - \kappa+\frac12 \right)}M_{\kappa, - \lambda}(2\varpi r) ,
\end{equation}
and
\begin{eqnarray}\label{N53}
M_{\pm\kappa ,\lambda}\left( {{e^{i\pi }}2\varpi r} \right) = (-1)^{\lambda + \frac12}M_{\mp\kappa ,\lambda}(2\varpi r), 
\end{eqnarray}
one finds
\begin{equation}\label{N51}
\tilde \Psi_1\left( r ,\varpi e^{i\pi } \right) = \tilde\Psi_1\left( r ,\varpi \right) ,
\end{equation}
and
\begin{equation}\label{N52}
{\tilde \Psi }_2\left( r,\varpi e^{i\pi } \right) = (-1)^{\left( \lambda+\frac12 \right)}\frac{\Gamma ( -2\lambda)}{\Gamma \left( - \lambda + \kappa+\frac12 \right)}M_{\kappa, \lambda}(2\varpi r) + (-1)^{\left( -\lambda+\frac12 \right)} \frac{\Gamma (2\lambda)}{\Gamma \left(\lambda + \kappa+\frac12 \right)}M_{\kappa, - \lambda}(2\varpi r) .
\end{equation}
It is worth noting that the master equation Eq.~\eqref{master-Whittaker} does not remain unchanged under the transformation $\varpi\to \varpi e^{i\pi}$.
In particular, $\kappa$ as the subscript of the Whittaker functions has flipped its sign twice, owing to Eqs.~\eqref{N53} and~\eqref{CoKappa}, where the second sign flip occurs to the master equation.
Also, $\lambda$ is invariant by Eq.~\eqref{CoLambda}.
We note that Eq.~\eqref{N51} confirms that $\tilde \Psi_1$ does not possess any branch cut.

To proceed, we consider the following three cases.

\subsection{The case $B=0$}\lb{subsec1}

In the literature, this case has been explored by several authors~\cite{agr-qnm-tail-31, agr-qnm-tail-32, agr-qnm-tail-33, agr-qnm-tail-34, agr-qnm-tail-37}, but somewhat distinct formalism has been utilized. 
The Wronskian Eq.~\eqref{defWronskian} now gives
\begin{equation}\label{NWrons}
W_1(\varpi ) = A( - 2\lambda)(2\varpi )\frac{\Gamma (2\lambda)}{\Gamma \left(\lambda - \kappa+\frac12 \right)} \equiv A(-2\lambda)(2\varpi)p_+,
\end{equation}
where one defines
\begin{eqnarray}\label{defPpm1}
p_+ &=& \frac{\Gamma (2\lambda)}{\Gamma \left(\lambda - \kappa+\frac12 \right)}  .
\end{eqnarray}
Also, the non-vanishing relation for the confluent hypergeometric functions~\cite{book-methods-mathematical-physics-05} has been employed to obtain
\begin{eqnarray}\label{Wrons}
%&&W\left\{M(a, b, z), z^{1-b}M(1+a-b, 2-b, z)\right\} = (1-b)z^{-b}e^z ,\nonumber \\
W\left\{ M_{\kappa, \lambda}(2\varpi r),  M_{\kappa, -\lambda}(2\varpi r) \right\} =  - (2\lambda)(2\varpi) .
\end{eqnarray}

To calcuate the discontinuity of the Wronskian across the branch cut, one makes use of Eq.~\eqref{N51} and~\eqref{N52} and the definition Eq.~\eqref{defWronskian} to find
\begin{equation}\label{NWronsFlip}
W_1\left( {\varpi {e^{i\pi }}} \right) = A ( -2\lambda)(2\varpi )\frac{\Gamma (2\lambda)}{\Gamma \left(\lambda + \kappa+\frac12 \right)}(-1)^{\left( -\lambda+\frac12 \right)} \equiv A(-2\lambda)(2\varpi)q_+,
\end{equation}
where
\begin{eqnarray}\label{defQpm1}
q_+ &=& \frac{\Gamma (2\lambda)}{\Gamma \left(\lambda + \kappa+\frac12 \right)}(-1)^{\left( -\lambda+\frac12 \right)} .
\end{eqnarray}
In deriving the above result, one should keep an eye on Eqs.~\eqref{N51} and~\eqref{N52}.
To be specific, the discontinuity embeded in $A$ does not lead to a factor $(-1)^{\left(-\lambda-\frac12\right)}$, while the factor $(-1)^{\left( -\lambda+\frac12 \right)}$ on the r.h.s. of Eq.~\eqref{N52} has come through.
Also, we note that Eq.~\eqref{NWronsFlip} cannot be obtained by simply flipping the sign of $\varpi$ and $\kappa$ in Eq.~\eqref{NWrons}.

By putting the pieces together, Eqs.~(\ref{N51}),~(\ref{N52}), and~\eqref{NWronsFlip}, one finds that the integrand given by Eq.~\eqref{defF} can be simplified to read
\begin{equation}\label{N38}
F_1(\varpi ) = \frac{1}{( - 2\lambda)(2\varpi )A^2}\tilde \Psi_1\left( r', \varpi  \right)\tilde F_1(\varpi )\tilde \Psi_1\left( r, \varpi \right) ,
\end{equation}
where 
\begin{equation}\label{tildeF}
\tilde F_1(\varpi ) =
(-1)^{2\lambda}\frac{\Gamma\left(-2\lambda\right)\Gamma\left(\lambda+\kappa+\frac12\right)}{\Gamma\left(-\lambda+\kappa+\frac12\right)\Gamma\left(2\lambda\right)}
-\frac{\Gamma\left(-2\lambda\right)\Gamma\left(\lambda+\frac12-\kappa\right)}{\Gamma\left(-\lambda-\kappa+\frac12\right)\Gamma\left(2\lambda\right)}  .
\end{equation}
Due to Eq.~(\ref{N51}), $\tilde \Psi_1\left( r' ,\varpi  \right)$ can be readily factorized out. 
The other term, $\tilde \Psi_2\left( r, \varpi  \right)$, can be simplified by separating the irrelevant contributions which possess identical discontinuity of the Wronskian. 
As a result, these terms cancel out in the subtraction, and from the remaining ones, a second common factor $\tilde \Psi_1\left( r, \varpi  \right)$ can be pulled out.

Now, we turn to discuss the asymptotic behavior of Green's function.
At a significant time scale $t \gg 1$, the integral receives contribution  primarily from the region where $\omega\to \pm \mu$ or $\varpi\to 0$, which implies that $\kappa\to \infty$.
The two $\tilde \Psi_1$ factors in Eq.~\eqref{N38} do not depend sensitively on the frequency $\omega$ when compared to the term Eq.~\eqref{tildeF}.
The latter contains exponential $\exp(i\pi\kappa)$ that oscillates significantly in the relevant frequency region.
This feature indicates the possibility of utilizing the {\it saddle point} approximation, where one estimates the result by considering the part where the oscillations evolve the slowest.

\begin{figure*}[htbp]
\centering
\includegraphics[scale=0.3]{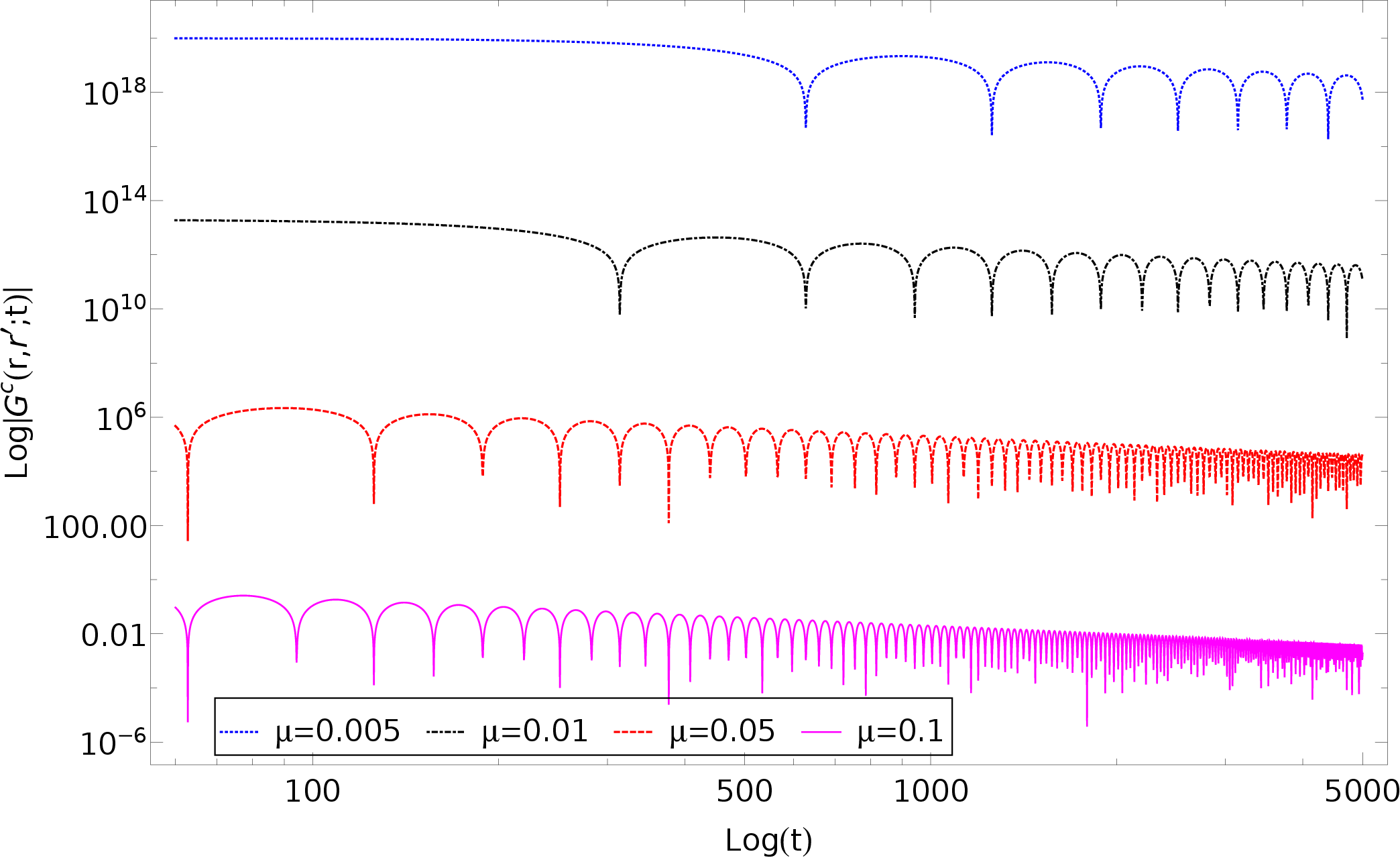}
\caption{The time-domain Green's function obtained by numerical integration.
The profiles are obtained by using parameters $M=1, \lambda=1$ for different masses $\mu=0.1, 0.05, 0.01$, and $0.005$.
For clarity, the data have been multiplied by different constants, $C= 10, 10^8, 10^{15}, 10^{22}$, respectively.}
\label{fig2tail}
\end{figure*}

\begin{figure*}[htbp]
\centering
\includegraphics[scale=0.3]{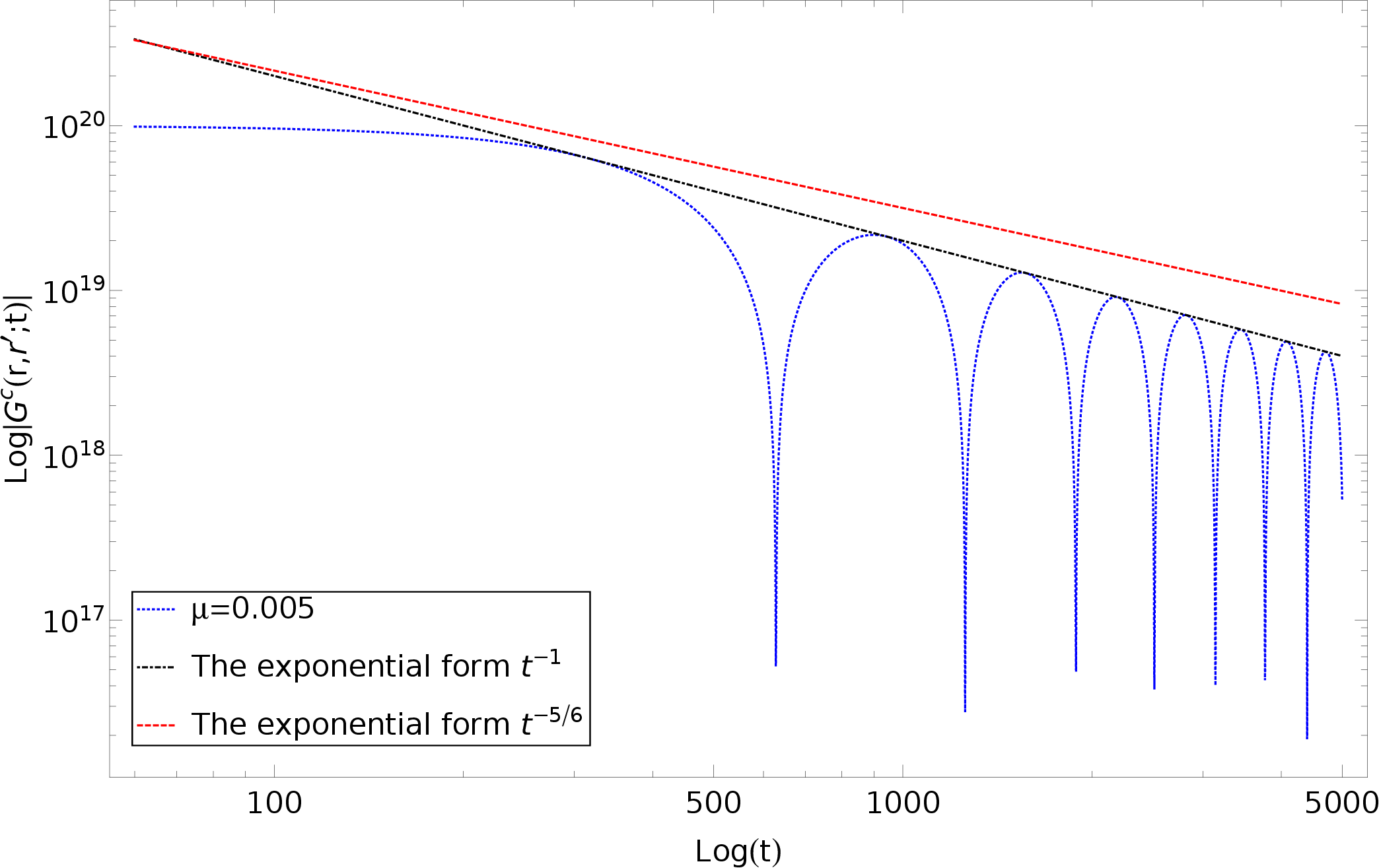}
\caption{The envelope of one of the curves of Fig.~\ref{fig2tail}, for $\mu=0.005$, is compared to the forms $t^{-1}$ and $t^{-5/6}$.}
\label{fig2comp}
\end{figure*}

Using the asymptotical forms of $\Gamma$ functions~\cite{book-methods-mathematical-physics-06}, one approximates Eq.~\eqref{tildeF} as
\begin{eqnarray}\label{N39}
\tilde F_1(\varpi ) &&\approx \frac{\Gamma\left(-2\lambda\right)}{\Gamma\left(2\lambda\right)}\kappa^{2\lambda}\left[(-1)^{2\lambda} - \frac{\eta e^{i\pi \kappa} + \gamma e^{-i\pi \kappa}}{\eta e^{-i\pi \kappa} + \gamma e^{i\pi \kappa}}\right] ,
%\tilde F_1(\varpi ) &&\approx \frac{\Gamma\left(-2\lambda\right)}{\Gamma\left(2\lambda\right)}\kappa^{2\lambda}\left[(-1)^{2\lambda} - \frac{\eta_+ e^{i\pi \kappa} + \eta_- e^{-i\pi \kappa}}{\eta_- e^{i\pi \kappa} + \eta_+ e^{-i\pi \kappa}}\right] ,
\end{eqnarray}
where
\begin{eqnarray}\label{etagamma}
\eta = -  e^{i\pi \left(\lambda-\frac12\right)},\ \ \
\gamma = e^{- i\pi \left(\lambda-\frac12\right)} .
%\eta_\pm  = \mp  e^{\pm i\pi \left(\lambda-\frac12\right)} .
\end{eqnarray}
At this point, it is rather attempting to rewrite the integrand as
\begin{equation}\label{N47}
\tilde F_1(\varpi )e^{-i\omega t} \approx \frac{\Gamma\left(-2\lambda\right)}{\Gamma\left(2\lambda\right)}\kappa^{2\lambda}{e^{i\phi }}  e^{i(2\pi\kappa-\omega t)},
\end{equation}
where the phase $\phi$ defined by
\begin{equation}\label{N48}
%{e^{i\phi }} =  - \frac{\eta_+  + \eta_- e^{-2i\pi \kappa}}{\eta_+ + \eta_- e^{2i\pi \kappa} } .
{e^{i\phi }} \stackrel{?}{=}  \frac{\eta  + \gamma e^{-2i\pi \kappa}}{\eta + \gamma e^{2i\pi \kappa} } 
\end{equation}
is hopefully a moderate function of $\omega$ as $\varpi\to 0$. 
Unfortunately, this is usually not viable for two reasons.
First, since $|\eta|=|\gamma|$, one cannot ignore the phase oscillation from this factor and move forward to employ the saddle point approximation described later in Sec.~\ref{subsec3}.
In fact, one can show geometrically that $\phi$, which is twice the inscribed angle, has the same magnitude of the central angle subtending the same arc, $2\pi\kappa$.
To be specific, we have
\begin{eqnarray}\label{PhiKappa}
\phi= -2\pi\kappa ,
\end{eqnarray}
where the initial phase shift $\arg\eta+1/2\arg(\gamma/\eta)$ cancels between the numerator and denominator,
and the negative sign before $2\kappa$ comes from the fact that the complex number $\gamma$ rotates clockwise and counter-clockwise on the numerator and denominator, respectively.
Eq.~\eqref{PhiKappa} implies that the intensive phase oscillations cancel out identically.
If the ratio $\gamma/\eta$ is a real number, Eq.~\eqref{N48} is exact.
In such case~\cite{agr-qnm-tail-32, agr-qnm-tail-33, agr-qnm-tail-40} for perturbations with insignificant mass, the asymptotic form of the tail is well-defined. 
To see this, one utilizes the specific form of $\Psi_1$, the Wronskian, and the expansion of Whittaker in $(2\varpi r)$ at the limit $\varpi \to 0$, and the integral simplifies to give 
\begin{eqnarray}\label{asympTail}
\int_{-\mu}^{\mu} e^{i\phi}e^{i(2\pi\kappa-\omega t)}d\omega \sim t^{-1}\sin(\mu t +\varphi_0) ,
\end{eqnarray}
where the phase $\varphi_0$ might be a minor function of $t$.
The cancelation of the phase and the resultant asymptotic form Eq.~\eqref{asympTail} can be readily verified by integrating numerically Eq.~\eqref{Fvarpi-contour} using Eqs.~\eqref{N38} and~\eqref{N39}.
The results are shown in Fig.~\ref{fig2tail} and~\ref{fig2comp}.
Nonlinear fits were carried out for the obtained profiles shown in Fig.~\ref{fig2tail}, and the results are given in Tab.~\ref{Tab1}, which are in reasonable consistency with the analytic form.
As shown explicitly in Fig.~\ref{fig2comp}, although numerically adjacent, this result is essentially different from the power-law $t^{-5/6}$ that was claimed previously by some authors.
The difference in the resulting exponential can be traced back to the (second) sign flip $\kappa\to -\kappa$ associated with the master equation due to Eq.~\eqref{CoKappa}, which was not taken into account in some literature.

\begin{table}[]
\caption{The extracted exponent from the envelope of the oscillations according to the form ${t^{ - \alpha }}\sin (\mu t)$.}\lb{Tab1}
\begin{ruledtabular}
\renewcommand\arraystretch{2}
\begin{tabular}{lllll}
$\mu $              & 0.005   & 0.01    & 0.05    & 0.1     \\
$\alpha $             & 1.019 & 1.009 & 1.058 & 1.054 \\
standard error & 0.041 & 0.029 & 0.024 & 0.022
\end{tabular}
\label{tab1}
\end{ruledtabular}
\end{table}

Second, when the ratio $\gamma/\eta$ is complex~\cite{agr-qnm-tail-31}, the module of Eq.~\eqref{N48} is not unit.
In fact, it becomes divergent when the denominator attains zero, at an increasing rate as $\kappa \to \infty$.
Although the Cauchy principal value of the integral Eq.~\eqref{Fvarpi-contour} may still exist, the result seems unmanageable.

\subsection{The case $A=0$}\lb{subsec2}

This case is similar to the previous subsection.
Instead of Eqs.~\eqref{NWrons},~\eqref{NWronsFlip},~\eqref{N38}, and~\eqref{tildeF}, we now have
\begin{equation}\label{NWrons2}
W_2(\varpi ) = B( + 2\lambda)(2\varpi )\frac{\Gamma (-2\lambda)}{\Gamma \left(-\lambda - \kappa+\frac12 \right)} \equiv B(+2\lambda)(2\varpi)p_-,\nb
\end{equation}
where
\begin{eqnarray}\label{defPpm2}
p_- &=& \frac{\Gamma (-2\lambda)}{\Gamma \left(-\lambda - \kappa+\frac12 \right)} .
\end{eqnarray}
\begin{equation}\label{NWronsFlip2}
W_2\left( {\varpi {e^{i\pi }}} \right) = B ( + 2\lambda)(2\varpi )\frac{\Gamma (-2\lambda)}{\Gamma \left(-\lambda + \kappa+\frac12 \right)}(-1)^{\left( \lambda+\frac12 \right)}\equiv B(+2\lambda)(2\varpi)q_- ,\nb
\end{equation}
where
\begin{eqnarray}\label{defQpm2}
q_- &=& \frac{\Gamma (-2\lambda)}{\Gamma \left(-\lambda + \kappa+\frac12 \right)}(-1)^{\left( \lambda+\frac12 \right)} ,
\end{eqnarray}
and the integrand Eq.~\eqref{defF} gives
\begin{equation}\label{N382}
F_2(\varpi ) = \frac{1}{( + 2\lambda)(2\varpi )B^2}\tilde \Psi_1\left( r', \varpi  \right)\tilde F_2(\varpi )\tilde \Psi_1\left( r, \varpi \right) , 
\end{equation}
where 
\begin{equation}\label{tildeF2}
\tilde F_2(\varpi ) =
(-1)^{-2\lambda}\frac{\Gamma\left(2\lambda\right)\Gamma\left(-\lambda+\kappa+\frac12\right)}{\Gamma\left(\lambda+\kappa+\frac12\right)\Gamma\left(-2\lambda\right)}
-\frac{\Gamma\left(2\lambda\right)\Gamma\left(-\lambda+\frac12-\kappa\right)}{\Gamma\left(\lambda-\kappa+\frac12\right)\Gamma\left(-2\lambda\right)}  .
\end{equation}
Again, the saddle point method cannot be applied to estimate the asymptotic behavior.

\subsection{The case $AB\ne 0$}\lb{subsec3}

By employing Eq.~\eqref{N36}, the Wronskian is found to be
\begin{eqnarray}\label{NWrons3}
W_3(\varpi ) = ( - 2\lambda)(2\varpi )\left[Ap_+-Bp_-\right] ,
\end{eqnarray}
where $p_\pm$ have been defined above by Eqs.~\eqref{defPpm1} and~\eqref{defPpm2}.

On the other hand of the branch cut, one finds
\begin{eqnarray}\label{NWronsFlip3}
W_3\left( {\varpi {e^{i\pi }}} \right) = ( - 2\lambda)(2\varpi )\left[Aq_+-Bq_-\right] ,\nb
\end{eqnarray}
where $q_\pm$ are given by Eqs.~\eqref{defQpm1} and~\eqref{defQpm2}.

The calculation of the integrand Eq.~\eqref{defF} is furnished by Eqs.~(\ref{N51}),~(\ref{N52}),~\eqref{NWrons3}, and~\eqref{NWronsFlip3}, and it demands a bit more tedious efforts.
It can be facilitated by the following algebraic relations
\begin{eqnarray}\label{NumRel1}
\frac12\left(Ap_++Bp_-\right)\tilde\Psi_1(\varpi)=\frac12\left[\left(Ap_+-Bp_-\right)\left(AM_{\kappa, \lambda}-BM_{\kappa, -\lambda}\right)\right]+AB\tilde\Psi_2(\varpi) , 
\end{eqnarray}
and
\begin{eqnarray}\label{NumRel2}
\frac12\left(Aq_++Bq_-\right)\tilde\Psi_1(\varpi)=\frac12\left[\left(Aq_+-Bq_-\right)\left(AM_{\kappa, \lambda}-BM_{\kappa, -\lambda}\right)\right]+AB\tilde\Psi_2(\varpi e^{i\pi}) . 
\end{eqnarray}
We observe that the terms in the brackets on the r.h.s. of Eqs.~\eqref{NumRel1} and~\eqref{NumRel2} possess the same discontinuity as their respective Wronskians.
As a result, the ratio to the Wronskian does not contain any discontinuity.
They are identical for both terms and readily canceled out when substituting into Eq.~\eqref{defF}.
Subsequently, one finds
\begin{equation}\label{N383}
F_3(\varpi ) = \frac{1}{( - 2\lambda)(4\varpi )AB}\tilde \Psi_1\left( r', \varpi  \right)\tilde F_3(\varpi )\tilde \Psi_1\left( r, \varpi \right) , 
\end{equation}
where 
\begin{equation}\label{tildeF3}
\tilde F_3(\varpi ) =
\frac{Aq_++Bq_-}{Aq_+-Bq_-}-\frac{Ap_++Bp_-}{Ap_+-Bp_-}  .
\end{equation}
It is noted that Eqs.~\eqref{N383} and~\eqref{tildeF3} are essentially Eqs.~(32-34) of Ref.~\cite{agr-qnm-tail-22}.
%However, if one assumes $B=0$ or $A=0$, Eq.~\eqref{tildeF3} vanishes identically, rather than falling back to Eqs.~\eqref{tildeF} or~\eqref{tildeF2}.
%Therefore, we point out that the particular cases discussed in Sec.~\ref{subsec1} and~\ref{subsec2} can not be inferred straightforwardly from the above result.
Although it might not be apparent from a first glimpse, Eqs.~\eqref{N38} and~\eqref{tildeF} can also be derived from Eq.~\eqref{tildeF3}.
For instance, Eq.~\eqref{tildeF3} seems to vanish by substituting $B=0$.
However, one may rewrite Eq.~\eqref{tildeF3} as
\begin{equation}\label{tildeF3B0}
\tilde F_3(\varpi ) =
2B\left[\frac{q_-}{Aq_+-Bq_-}-\frac{p_-}{Ap_+-Bp_-}\right]  ,\nb
\end{equation}
which, after the factor $2B$ partially cancels out the term $AB$ in the demoninator in Eq.~\eqref{tildeF3}, gives Eq.~\eqref{N38} by assuming $B=0$.

As pointed out by Koyama and Tomimatsu~\cite{agr-qnm-tail-20, agr-qnm-tail-21, agr-qnm-tail-22}, one might employ the saddle point approximation. 
From Eq.~\eqref{tildeF3}, one separates a relevant factor that oscillates most drastically and uses it to estimate the integration. 
One must justify that it only oscillates moderately for the remaining term, even though its apparent form might be rather sophisticated.
By using the asymptotic forms of $\Gamma$ functions~\cite{book-methods-mathematical-physics-06} at the limit $\kappa\to \infty$, it is not difficult to show that the first term of the r.h.s. of Eq.~\eqref{tildeF3} is irrelevant.
For the second term, it gives
\begin{equation}\label{tildeF32}
\frac{\eta_+e^{i\pi\kappa} +\gamma_+e^{-i\pi\kappa}}{\eta_-e^{-i\pi\kappa} +\gamma_-e^{i\pi\kappa}},  
\end{equation}
where
\begin{eqnarray}
  \eta_{\pm}&=&\Gamma(2\lambda)A\kappa ^{-\lambda}e^{-i\pi\lambda}\pm \Gamma(-2\lambda)B\kappa ^{\lambda}e^{i\pi\lambda} ,\nb \\
  \gamma_{\pm}&=&\Gamma(2\lambda)A\kappa ^{-\lambda}e^{i\pi\lambda}\pm \Gamma(-2\lambda)B\kappa ^{\lambda}e^{-i\pi\lambda} .
\end{eqnarray}

Now, if one can show that $|\eta_\pm| > |\gamma_\pm|$ for $\omega > 0$, then the r.h.s. of Eq.~\eqref{tildeF32} can be rewrite as
\begin{equation}\label{FN48}
\tilde F_3(\varpi )e^{-i\omega t} \sim {e^{i\phi }}e^{i(2\pi\kappa-\omega t)},
\end{equation}
where
\begin{equation}\label{iPhiN48}
{e^{i\phi }} =  \frac{\eta_+  + \gamma_+ e^{-2i\pi \kappa}}{\eta_- + \gamma_- e^{2i\pi \kappa} } .
\end{equation}
To be specific, we have
\begin{eqnarray}\label{iPhiN48}
\phi =&& \ \left(\arg\eta_+-\arg\eta_-\right) \nb\\
&&+\arctan\frac{|\gamma_+|\sin(\arg\gamma_+-\arg\eta_+-2\pi\kappa)}{|\eta_+|+|\gamma_+|\cos(\arg\gamma_+-\arg\eta_+-2\pi\kappa)} \nb\\
&&-\arctan\frac{|\gamma_-|\sin(\arg\gamma_--\arg\eta_-+2\pi\kappa)}{|\eta_-|+|\gamma_-|\cos(\arg\gamma_--\arg\eta_-+2\pi\kappa)} .
\end{eqnarray}
When $\kappa\to \infty$, it is noted that the argument $\phi$ also oscillates violently by the same period as $2\pi\kappa$.
It varies within the range of $2\pi$, and as discussed below, its effect will not completely compensate for the latter.
On the other hand, if for $\omega < 0$, we have $|\eta_\pm| < |\gamma_\pm|$, and the same conclusion holds.

The saddle point is identified at the vanishing rate of change for the phase, namely, 
\begin{equation}\label{N49}
\frac{d}{{d\omega }}\left( {2\pi \kappa +\phi - \omega t} \right) = 0 ,
\end{equation}
from which one finds
\begin{equation}\label{N50}
{\varpi _0} = \sqrt { \mu^2 - {\omega_0 ^2}}  \simeq {\left( {\frac{{C \mu^3 M}}{t}} \right)^{\frac{1}{3}}} ,
\end{equation}
where
\begin{equation}
C = 2\pi +\left.\frac{d\phi}{d\kappa}\right|_{\varpi=\varpi_0} .\nb
\end{equation}
We note that complete cancelation does not occur because
\begin{equation}
%\left.\frac{d\phi}{d\kappa}\right|_{\varpi=\varpi_0} 
%\sim (-2\pi)\left.\frac{\frac{|\gamma_+|\cos(2\pi\kappa)}{|\eta_+|+|\gamma_+|\cos(2\pi\kappa)}+\left(\frac{|\gamma_+|\sin(2\pi\kappa)}{|\eta_+|+|\gamma_+|\cos(2\pi\kappa)} \right)^2}{1+ \left(\frac{|\gamma_+|\sin(2\pi\kappa)}{|\eta_+|+|\gamma_+|\cos(2\pi\kappa)} \right)^2}\right|_{\varpi=\varpi_0}
%\sim (-2\pi)\left.\frac{|\eta_+||\gamma_+|\cos(2\pi\kappa)+|\gamma_+|^2}{|\eta_+|^2+2|\eta_+||\gamma_+|\cos(2\pi\kappa)+|\gamma_+|^2}\right|_{\varpi=\varpi_0}
%+ \mathrm{c.c.}
\frac{d\phi}{d\kappa} 
\sim (-2\pi)\frac{|\eta_+||\gamma_+|\cos(\arg\gamma_+-\arg\eta_+-2\pi\kappa)+|\gamma_+|^2}{|\eta_+|^2+2|\eta_+||\gamma_+|\cos(\arg\gamma_+-\arg\eta_+-2\pi\kappa)+|\gamma_+|^2}
+ \mathrm{c.c.}
> -2\pi ,\nb
\end{equation}
where one estimates the dominant contribution from the exponential dependence on $\kappa$,
and the counter term ``$\mathrm{c.c.}$'' is obtained by the replacements $\eta_+\to\eta_-, \gamma_+\to\gamma_-$ and $\kappa\to -\kappa$.

By further evaluating the derivatives
\begin{eqnarray}\label{N61}
K_0 &=& {\left. {(2\pi \kappa +\phi  - \omega t )} \right|_{\omega  = {\omega _0}}} \simeq 2\pi\kappa(\omega_0)+\phi(\omega_0)-\mu t,\nonumber\\
K_2 &=& {\left. {\frac{{{d^2}}}{{d{\omega ^2}}}\left( {2\pi \kappa +\phi - \omega t} \right)} \right|_{\omega  = {\omega _0}}} \simeq \frac{{3\mu t}}{{{{C}^{2/3}}{{\left( \frac{\mu  ^3 M }{t}\right)}^{2/3}}}} ,
\end{eqnarray}
one finds the desired asymptotic form for the time-domain Green's function Eq.~\eqref{Fvarpi-contour}
\begin{equation}\label{N50}
{G^C}\left(r, r'; t\right) \sim e^{iK_0}\int e^{i\frac12  K_2\rho^2} d\rho \sim {(\mu t)^{ - 5/6}}{(\mu M)^{1/3}}\mu \sin (\mu t +\varphi_0) ,
\end{equation}
where the direction of the steepest descent is alone $\omega ~ \pm \exp(i\pi/4)$.
and the phase shift $\varphi_0$ is governed by $\phi (\omega _0)$ and $\kappa(\omega_0)$, with some minor dependence in $t$.
The resultant temporal dependence of the envelope is, therefore, $t^{ - 5/6}$, as explored by many authors.

If $|\eta_\pm| = |\gamma_\pm|$ and $\lambda$ is a real number, by using the same arguments we have $\phi = -2\pi\kappa$.
Therefore, the strong oscillation as $\varpi\to 0$ cancels out, and subsequently, at the small mass limit, we also arrive at a late-time tail of the form $t^{ - 1} $.

\section{Further discussions and concluding remarks} \label{section4}

In this study, we gave a detailed account of the derivations of the late-time tails of massive perturbations in spherically symmetric black hole metrics.
A few aspects have been revisited regarding the theoretical framework initially pioneered by Hod and Piran~\cite{agr-qnm-tail-16}, and Koyama and Tomimatsu~\cite{agr-qnm-tail-22}.

The Green's function Eq.~\eqref{defGreen} in the context of black hole perturbation theory is strongly reminiscent of that in classical electrodynamics.
In particular, if the effective potential is suppressed, the formalism readily falls back to that of the Lorentz gauge Maxwell equations for the scalar and vector potentials in free space~\cite{book-classical-electrodynamics-Jackson}.  
Although different types of Green's functions obey the same equation, it is the {\it causality}, in terms of specific boundary conditions, that fixes the remaining ``liberty''. 
In the language of contour integration, the difference between the retarded and advanced Green's functions correspond to distinct choices in shifting the mass poles around the real frequency axis~\cite{book-classical-electrodynamics-Jackson}.
The above notion continues to be the same in quantum field theory, except that the role of the free field Green's functions is replaced by the propagators~\cite{book-qft-Peskin-Schroeder}.
The current approach constructs the Green's function using Eq.~\eqref{GFomega}, where the retarded nature of the resulting formula is planted through the asymptotic behavior of the solution for the homogeneous equation.
Therefore, in principle, there is no more freedom regarding the locations of the QNM poles and branch cuts.
In other words, these properties are well-defined.
Indeed, the whereabouts of the poles and cuts can be correctly ``guessed'' using heuristic arguments based on the definition of retarded Green's function and Jordan's lemma, as has been done in most literature.
Nonetheless, it is meaningful to ascertain the mathematical consistency of the theory through an explicit analysis of the analytic properties of the waveforms.
Moreover, since the derived master equation is an approximation, we understand that such a reassessment, as carried out in the first part of Sec.~\ref{section3}, is relevant.  

In deriving the master equation Eq.~\eqref{masterEq-Whittaker}, we have ignored the first-order derivative term 
\begin{eqnarray}
\frac{M+M'}{r^2}\frac{\partial \tilde{\Psi}}{\partial r},  \nb
\end{eqnarray}
which is of $O(1/r^2)$.
On the other hand, for the second-order derivative, such terms were kept on the r.h.s. of Eq.~\eqref{effU}.
This can be justified as follows.
First, the resulting late-time tail was independent of the value of $\lambda$ and therefore readily applied to the case where the term vanishes by assuming $\lambda=\frac14$.
Indeed, the specific form of $\lambda$ given by Eq.~\eqref{CoLambda} is not essential, as one only needs the property that is invariant under the transformation $\varpi\to \varpi e^{i\pi}$.
Second, the relevant region of the saddle point, the frequency, essentially vanishes and $\kappa\to \infty$, which again ensures that the last term's effect is relatively insignificant.
It is also consistent with the obtianed asymptotic behavior of $\tilde\Psi_2$ at large $r$, the ratio of the two terms is largely determined by ${\tilde{\Psi}_2}'/\tilde{\Psi}_2 \sim \varpi \to 0$.
Subsequently, many authors have interpreted such an approximation as that the late-time tail is due to the backscattering from the effective potential at spatial infinity, reminiscent of the case of massless perturbations~\cite{agr-qnm-tail-06}.

In~\cite{agr-qnm-tail-22}, the authors elaborate a physical interpretation for the condition when the asymptotic tail behaves universally as $t^{-5/6}$.
It states that the fraction of the in-going wave in $\Psi_1$ should be more significant than that of the out-going wave.
Indeed, the forms of $\tilde\Psi_1$ discussed in Sec.~\ref{subsec1} and~\ref{subsec2} possess both in-going and out-going components.
Moreover, for the specific cases discussed in this study, it is not difficult to show that the out-going wave becomes the dominant one.
To be specific, this occurs since $\kappa$ is significant in our case. 
Besides, due to the asymptotic form of the confluent hypergeometric function $M(\pm\lambda-\kappa+1/2, 1\pm 2\lambda, x)$, the coefficient of the out-going wave becomes more pronounced.
The latter can be readily identified using the phase given by Eq.~\eqref{argVP}, as a result of the boundary constraint for $\tilde\Psi_2$.
Therefore, the findings of the present work are consistent with the existing results.

\section*{Acknowledgments}
This work is supported by the National Natural Science Foundation of China (NNSFC) under contract No. 11805166.
We also gratefully acknowledge the financial support from
Funda\c{c}\~ao de Amparo \`a Pesquisa do Estado de S\~ao Paulo (FAPESP),
Funda\c{c}\~ao de Amparo \`a Pesquisa do Estado do Rio de Janeiro (FAPERJ),
Conselho Nacional de Desenvolvimento Cient\'{\i}fico e Tecnol\'ogico (CNPq),
Coordena\c{c}\~ao de Aperfei\c{c}oamento de Pessoal de N\'ivel Superior (CAPES),
A part of this work was developed under the project Institutos Nacionais de Ci\^{e}ncias e Tecnologia - F\'isica Nuclear e Aplica\c{c}\~{o}es (INCT/FNA) Proc. No. 464898/2014-5.
This research is also supported by the Center for Scientific Computing (NCC/GridUNESP) of the S\~ao Paulo State University (UNESP).

\appendix

\section{Analytic properties of the in-going wave $\tilde\Psi_1$}\label{indxA}

In this Appendix, we elaborate further on the singularity of the in-going wave $\tilde\Psi_1$.
One may argue that if the effective potential vanishes in the vicinity of the horizon at least as fast as an exponential form, the waveform does not contain any branch cut.
As shown in~\cite{agr-qnm-tail-06}, if the potential decrease faster than an exponential form, there will be no singularity.
The waveform contains a pole on the negative imaginary axis for the marginal case of an exponential potential, but there is no branch cut.
However, since we are interested in the late-time tail, a branch cut stemming from the real axis plays a more significant role, and a non-oscillating pole is irrelevant.
This is because a power law decay, dictated by the cut, overwhelms an exponential one, governed by a pole. 
Now, for a generic black hole metric, one can show that the effective potential indeed decreases asymptotically by an exponential form.
Roughly speaking, this is because the effective potential $V(r)$ attains zero linearly, meanwhile the tortoise coordinate implies $r_*\propto \ln(r-r_h)$.
Based on the above arguments, it is physically plausible that the in-going wave does not contain a branch cut, and as a result, the asymptotic behavior of the late-time tail is mainly governed by the singularities in the out-going wave.
For instance, Eq.~(17) of Ref.~\cite{agr-qnm-tail-20} gives a general form of retarded Green's function, and the asymptotic tail is governed mainly by the singularities embedded in $\alpha$ and $\beta$, rather than the particular form of the in-going wave.
A more specific example for massless perturbations is given by Eq.~(4.12) Ref.~\cite{agr-qnm-tail-06}, where it is evident that the specific form of the in-going wave does not play a significant role in the temporal dependence given later in Eq.~(4.17).
We note that the above examples' conclusions are rather general, not restricted to the specific black hole metrics and approximations.

In practice, the analytic properties of the in-going wave are often taken as a requirement, primarily when it is obtained from an approximated potential and substantially distorted.
In the present work, for example, the master equation Eq.~\eqref{masterEq} is derived as a proximation to the potential at spatial infinity.
Therefore, it is not expected to precisely reproduce the black hole metric in the vicinity of the horizon.
Nonetheless, according to the above discussions, it will not present a serious problem as long as the waveform $\tilde\Psi_1$ does not contain any branch cut.
In what follows, we show that this is indeed the case.
Concerning the discussions around Eq.~\eqref{argVP}, for the in-going wave, the boundary condition dictates a relation similar to that presented in the main text.
Again, this implies a jump in the phase of $\varpi$, namely, $\varpi\to\varpi e^{i\pi}$ as one traverses the line segment between $\pm \mu$ on the real axis of the frequency $\omega$. 
However, owing to Eq.~\eqref{N51}, this will not cause a difference in value (either the derivatives) of $\tilde \Psi_1$ between the two sides on the branch.
In other words, even though $\varpi$ as a function of $\omega$ inevitably possesses a branch cut, $\tilde \Psi_1$ as a function of $\omega$ is manifestly analytic in the relevant domain.
As discussed above, this was motivated and intentionally chosen.

\bibliographystyle{JHEP}
\bibliography{references_qian}

\end{document}